# Delayed Rejection Variational Monte Carlo

Dario Bressanini[1], Gabriele Morosi[2], and Silvia Tarasco[3]

*Dipartimento di Scienze Chimiche ed Ambientali, Universita' dell'Insubria, Sede di Como, via Lucini 3, 22100 Como, Italy*

Antonietta Mira[4]

*Dipartimento di Economia, Universita' dell'Insubria, Sede di Varese, via Ravasi 2, 21100 Varese, Italy*

**Abstract**

A new acceleration algorithm to address the problem of multiple time scales in variational Monte Carlo simulations is presented. After a first attempted move has been rejected, the delayed rejection algorithm attempts a second move with a smaller time step, so that even moves of the core electrons can be accepted. Results on Be and Ne atoms as test cases are presented. Correlation time and both average accepted displacement and acceptance ratio as a function of the distance from the nucleus evidence the efficiency of the proposed algorithm in dealing with the multiple time scales problem.

---

[1] Electronic address: Dario.Bressanini@uninsubria.it

[2] Electronic address: Gabriele.Morosi@uninsubria.it

[3] Electronic address: Silvia.Tarasco@uninsubria.it

[4] Electronic address: Antonietta.Mira@uninsubria.it

# INTRODUCTION

Variational Monte Carlo (VMC) has become an important technique in quantum chemistry. When analytical integration is not available, it allows the computation of expectation values of an arbitrary trial wave function[1] with no restriction on its functional complexity. The trial wave function can include explicit two-body and higher-order correlation terms, allowing a better description of many body interactions and thus higher accuracy. The optimization of the variational parameters can be done by minimizing the energy[2-5], the energy variance[6-9] or the mean absolute deviation of the local energy[10]. Nevertheless, the main problem of any stochastic method is the need of reducing the statistical uncertainty on the calculated quantities. For this reason large systems present a computational challenge: in particular, Monte Carlo, as all total energy methods, suffers of scaling problems: an increase of the size of the system gives rise to an explosion in the computational cost, proportional to (often) large powers of the system size. This large-power polynomial scaling is surely preferable to the computational exponential dependency on the system size of NP problems; nevertheless, it prevents the treatment of many physically interesting large systems.

Another drawback is that systems containing atoms of large atomic number (Z) would require different time steps in order to efficiently sample both the regions close and far away from the nuclei: this problem is referred to as the multiple time scales problem. Core electrons require a smaller time step than valence electrons. This causes an intrinsic algorithmic inefficiency since the standard Metropolis algorithm assigns the same time step to all the electrons. Thus sampling in the region close to the nucleus is the bottleneck of VMC simulations. Although Monte Carlo methods scale well with N, they show a poor scaling with the atomic number Z; CPU time is estimated to scale with $Z^{5.5}$ or $Z^{6.5}$.

Analysing the standard Monte Carlo algorithm, Bressanini and Reynolds[11] showed that the optimal move size is a trade-off between the best move size for electrons far from the nucleus (i.e.

valence electrons), which needs to be large since the accessible region of configuration space is very large, and the best move size for the electrons close to the nucleus (i.e. core electrons). These latter moves must be small, since the relevant region of configuration space is quite limited, and also because the wave function changes rapidly near the nucleus, meaning that large moves would cause a high rejection rate.

Acceleration algorithms have been suggested to cope with the multiple time scale problem. Belohrec et al.[12] proposed the "split-tau" technique, that is they used different time scales for different shells, dividing the electrons into shells on the basis of their distance from the nucleus. Trying to assign a different time step (and so a different time scale) to different electrons does not work. Given a symmetric or antisymmetric wave function, two identical particles (here like-spin electrons) can exchange positions without changing the probability of the configuration. Thus, assigning larger time steps to electrons starting out in the valence region at the beginning of the simulation would not accomplish the goal, since ultimately such electrons exchange their positions with inner electrons, with no energy penalty. Once this happens, the electrons take inappropriate step sizes and detailed balance is no more satisfied. In formulating their "split-tau" technique, Belohrec et al.[12] had to assume that the exchange between shells is negligible. However, Sun et al.[13] results do not support this hypothesis, showing that it is true only for very small time steps, that is for very inefficient simulations.

Umrigar[14] proposed the factorization of the transition matrix in radial and angular parts. The overall move is large, but the nearer the electron is to the nucleus, the more its angular move is increased reducing the radial move and so increasing the acceptance probability. Sun et al.[13] showed that the addition of a simplified quadratic approximation to the wave function to sample the electron displacement automatically reduces the step size of the core electrons defined by the linear approximation, i.e., by the quantum force. Mella et al.[15] used a transition matrix in which the time step depends on the actual position of the electron: a suitable choice allows the control of the effect of the quantum force near the nucleus and near the nodal surfaces. Bressanini and Reynolds[11] showed that, after partitioning the space into equivalent subspaces, it is possible to choose independent sampling times for core and valence electrons.

Nevertheless, none of the above mentioned approaches is general. Some of these solutions are impossible to generalize, while others are very difficult to implement.

Here we propose a simple algorithm, easy to implement, completely general and that allows to sensibly improve the sampling.

## OVERVIEW OF VMC

Since very detailed descriptions of VMC are available elsewhere[16], we only give a short resume. VMC allows to sample a distribution proportional to $\Psi_T^2(\mathbf{R})$, where $\Psi_T(\mathbf{R})$ is a trial wave function. From such a distribution expectation values of non-differential operators can be obtained simply by:

$$\left\langle \hat{O} \right\rangle = \frac{\int \Psi_T^2(\mathbf{R})\hat{O}(\mathbf{R})d\mathbf{R}}{\int \Psi_T^2(\mathbf{R})d\mathbf{R}} \cong \frac{1}{N}\sum_{i=1}^{N} \hat{O}(\mathbf{R}_i) \qquad (1)$$

Differential operators can also be simply treated, by writing

$$\left\langle \hat{O} \right\rangle = \frac{\int \Psi_T^2(\mathbf{R})\frac{\hat{O}\Psi_T(\mathbf{R})}{\Psi_T(\mathbf{R})}d\mathbf{R}}{\int \Psi_T^2(\mathbf{R})d\mathbf{R}} \cong \frac{1}{N}\sum_{i=1}^{N} \frac{\hat{O}\Psi_T(\mathbf{R}_i)}{\Psi_T(\mathbf{R}_i)} \qquad (2)$$

The problem reduces to sample efficiently a distribution proportional to $\Psi_T^2(\mathbf{R})$. A set of walkers at positions $\mathbf{R}_i^{'}$ is displaced to new positions $\mathbf{R}_i$ by moving each walker. In the standard Metropolis algorithm a step is generated by "box sampling", that is $\mathbf{R} = \mathbf{R}^{'} + \vec{x}\Delta$, with $\Delta$ the step size and $\vec{x}$ a 3-N dimensional vector of uniformly distributed random numbers $\vec{x} \in [-0.5, +0.5]$. This move is followed by the classical Metropolis accept/reject step, in which $\left(\Psi_T(\mathbf{R})/\Psi_T(\mathbf{R}^{'})\right)^2$ is compared to a uniformly distributed random number between zero and one. The move is accepted only if the squared ratio of trial functions exceeds the random number, otherwise the old position is retained. This is one step of the Markov chain. Under very general conditions, this chain results in an asymptotic equilibrium distribution proportional to $\Psi_T^2(\mathbf{R})$.

The sampling can be improved using the Langevin sampling algorithm. This scheme is a generalization of the Metropolis sampling in which a Langevin equation $\mathbf{R} = \mathbf{R}' + Dt F(\mathbf{R}') + \bar{c}$, containing drift and diffusion (i.e. a "quantum" force F(**R'**) and a white noise $\bar{c}$, a Gaussian random variable with a mean value of zero and a standard deviation $2D\tau$), is employed. The quantum force depends on the position, but the overall attempted move is still determined by the time step. Once again, the use of a single time step for all electrons implies a kind of negotiation between valence and core electrons.

## DELAYED REJECTION MONTE CARLO

Tuning the time step of the Metropolis algorithm is not an easy task. As previously outlined, there is a trade-off between the time-step and the acceptance ratio of the corresponding proposed move. Furthermore, if we move one electron at a time (local moves), the "optimal" time step for each move depends on the distance of the electron from the nucleus: the closer a particle is to the nucleus (core region), the smaller the time step should be.

The origin of our proposal is the simple observation that, using the same time step, core-electron moves are rejected more than valence-electron moves. The previously proposed acceleration techniques tried to prevent this rejection, the delayed rejection algorithm, instead, uses this information and tries a second stage proposals to improve the sampling by the Metropolis and Langevin algorithms.

## *The delayed rejection strategy*

In a generalized Metropolis algorithm, one samples $p(\mathbf{R}) = \Psi_T(\mathbf{R})^2$ by constructing a Markov chain. Given the current position of the chain at the n-th step, $\mathbf{R}^{(n)} = \mathbf{R}'$, a candidate move $\mathbf{R}_1$ during a time step $t_1$ is generated by a given transition probability $T_1(\mathbf{R}' \to \mathbf{R}_1; t_1)$. The proposed move is accepted with probability

$$P_1(\mathbf{R}', \mathbf{R}_1) = \min\left[1, \frac{p(\mathbf{R}_1)T_1(\mathbf{R}_1 \to \mathbf{R}'; t_1)}{p(\mathbf{R}')T_1(\mathbf{R}' \to \mathbf{R}_1; t_1)}\right] \tag{3}$$

so that detailed balance with respect to $p(\mathbf{R})$, and thus stationarity, is preserved.

If the move is accepted, the simulation time is advanced and the chain position is updated: $\mathbf{R}^{(n+1)} = \mathbf{R}_1$.

So far the updating mechanism of the Markov chain is just like the one used for a regular Metropolis algorithm. In the delayed rejection algorithm the difference is in what happens upon rejection of the candidate move. In the Metropolis scheme, upon rejection the simulation time is advanced and the current position is retained: $\mathbf{R}^{(n+1)} = \mathbf{R}^{(n)} = \mathbf{R}'$. Although remaining in the current state contributes to preserve the stationary distribution through detailed balance, intuitively it increases autocorrelation in the realized chain and thus reduces the efficiency of the resulting estimators. Substance is given to this intuition by a result stated, and proved for the case of a finite state space, by Peskun[17]; a proof for general state spaces was given by Tierney[18]. Given two Markov chains with stationary distribution π, $T_1$ and $T_2$ being the corresponding transition matrices, $T_1$ is more efficient than $T_2$ (in the sense of reducing the asymptotic variance of the resulting estimators and thus the autocorrelation time, for any stochastic variable X) if

$$T_1(\mathbf{R}' \to \mathbf{R}) \geq T_2(\mathbf{R}' \to \mathbf{R}) \quad \forall \mathbf{R} \neq \mathbf{R}' \tag{4}$$

In other words, the higher the probability of moving away from the current position, the better the efficiency. Following this intuition, in the delayed rejection strategy[19], upon rejection of a first stage candidate move **R₁**, a second stage candidate move, **R₂**, is proposed by generating it from a different transition probability $T_2(\mathbf{R}' \to \mathbf{R}_2; t_2)$. Now the transition probability $T_2(\mathbf{R}' \to \mathbf{R}_2; t_2)$ is allowed to depend on the rejected value at the first stage. To cope with the multiple time step problem, a time step $t_2$ can now be chosen shorter than the one previously used. In order to maintain the reversibility condition and to preserve the stationary distribution, the acceptance probability of the new candidate move must be adjusted as

$$P_2(\mathbf{R}', \mathbf{R}_1, \mathbf{R}_2) = \min\left[1, \frac{p(\mathbf{R}_2)T_1(\mathbf{R}_2 \to \mathbf{R}_1; t_1)(1 - P_1(\mathbf{R}_2, \mathbf{R}_1))T_2(\mathbf{R}_2 \to \mathbf{R}'; t_2)}{p(\mathbf{R}')T_1(\mathbf{R}' \to \mathbf{R}_1; t_1)(1 - P_1(\mathbf{R}', \mathbf{R}_1))T_2(\mathbf{R}' \to \mathbf{R}_2; t_2)}\right] \quad (5)$$

to preserve the detailed balance condition. If $\mathbf{R}_2$ is accepted, we set $\mathbf{R}^{(n+1)} = \mathbf{R}_2$. Otherwise the delayed rejection process can either be interrupted by setting $\mathbf{R}^{(n+1)} = \mathbf{R}^{(n)} = \mathbf{R}'$, or continued with higher stage proposals using an iterative formula for the acceptance probability[20]. Since the acceptance probabilities preserve detailed balance separately at each stage, hybrid strategies can also be considered: upon rejection a coin is tossed and depending on the result the delayed rejection process is either continued or interrupted. The second attempted move reduces the overall probability of remaining in the current state. It can be proved[19] that an algorithm with delayed rejection dominates, in the Peskun sense, the corresponding standard algorithm. This is true for Metropolis as well as for Langevin algorithms. The autocorrelation time for any stochastic variable X is reduced by adding one or more delayed rejection steps. Taking different transition probabilities corresponding, for example, to two different time steps $t_1 \rangle t_2$ allows moving particles far from the nucleus at the first stage and particles in the core at the second stage. Both moves can either be local (one electron at a time) or global (all electrons at once). In a similar way the delayed rejection strategy can be used to combine global (first stage) with local moves (second stage): again global moves are less likely to be accepted, but faster to perform from a computational point of view. The usefulness of the delayed rejection algorithm depends on whether the obtained reduction in variance compensated for the additional computational cost.

## RESULTS AND DISCUSSION

Our main goal introducing the delayed rejection algorithm in VMC is to improve the efficiency of the method, allowing the electrons to move both near the core and far from it. So, as test cases we chose Be and Ne atoms to compare the effect of different Z values. In particular Ne was studied by Sun *et al.*[21], Mella *et al.*[15], and Bressanini and Reynolds [11], so our results can be compared with those of different acceleration algorithms. The efficiency is measured by the asymptotic variance of the estimator of the quantity of interest (typically the local energy E(**R**)), with respect to a distribution $p(\mathbf{R})$ known up to a normalizing constant. Since the estimator is the average of the sampled values $E_i$ along the Markov chain, its asymptotic variance is the sum of the autocorrelations of $E_i$ along such path. So, the autocorrelation time of the local energy can be considered a natural measure of efficiency. It is related to the time one needs to obtain decorrelated measures of the observable and the smaller it is, the more efficient is the algorithm. The autocorrelation time depends both on the sampling inefficiency of the algorithm and on the trial wave function through the fluctuations of the local energy. Thus, in order to make comparisons with different sampling methods, one must use the same trial wave function. In this work a simple SCF wave function multiplied by an electron-electron Jastrow factor was chosen. The Clementi and Roetti basis set[22] was employed for Be atom, while for Ne atom a DZ basis set was optimized. Simulations were performed with different time steps, trying to minimize the energy autocorrelation time. The autocorrelation time is a "macroscopic" measure[21] of the simulation efficiency, it provides information on long term, accumulated effects. To investigate the problem of the multiple time scales, a "microscopic" analysis is more informative. The space around the nucleus was divided in spherical shells and the acceptance ratio and the mean accepted displacement in each shell was estimated. First of all, the electron displacement was evaluated for the standard

Metropolis algorithm, in which the move is accepted or rejected only when all electrons have moved to new positions. Since in the delayed rejection algorithm each electron moves independently of the others, also the results obtained with a standard Metropolis algorithm, but moving one electron at a time, were examined. Then this analysis was repeated using the standard Langevin algorithm, again moving all electrons at once and one electron at a time.

The delayed rejection algorithm was implemented within the framework of both Metropolis and Langevin algorithms. A comparison between the results obtained by our algorithm and the standard ones, both moving all electrons at once and one electron at a time is now presented.

Let us begin discussing the autocorrelation time of the local energy, the "macroscopic" measure of the efficiency of a sampling algorithm The correlation times for simulations using the different algorithms are reported in Table 1 for Be and in Table 2 for Ne. The time steps are those that minimize the correlation time. For comparable correlation times the Ne time steps are shorter than for Be, the more compact is the atomic core the shorter must be the move. Moving one electron at a time allows longer time steps, due to the possibility of electrons to move independently, and is more efficient than moving all electrons at once, a well known fact. The Langevin sampling is more efficient than the Metropolis one as a consequence of the effect of the quantum force on the electron move. For single electron moves the best time steps are equal for Langevin and Metropolis sampling, but the Langevin correlation time is smaller. The delayed rejection algorithm further on improves the efficiency both of Metropolis and Langevin Monte Carlo, for Ne the correlation time is halved. The first time step is more than twice the corresponding value in absence of delayed rejection, improving the sampling of the valence space. The time step for the second move is one order of magnitude smaller than the first one, so also the core space is efficiently sampled. The decrease of the second time step with respect to the first one is larger for Ne than for Be, due to the presence of a more compact atomic core that requires shorter time steps to sample the core region.

In the following "microscopic" analysis we will discuss only the Ne case, as the results for Be and Ne are similar, but the effect of the delayed rejection algorithm is more evident for Ne. The number of attempted and accepted moves for the delayed rejection Langevin simulation with $\tau_1=0.03$ hartree$^{-1}$ and $\tau_2=0.007$ hartree$^{-1}$ as function of the distance from the nucleus are shown in Fig. 1. The attempted move distribution at the first step obviously reproduces the electron density of the Ne atom with its shell structure. For r>1.4 bohr nearly all attempted moves are accepted, but for shorter distances the number of accepted moves goes quickly to zero. The second step allows to recover a substantial fraction of the rejected moves. A similar plot for delayed rejection Metropolis evidences a minor number of accepted moves at the first step in the valence region, but a larger number at the second step in the core space. This result prompted us to perform the first step with the Langevin algorithm and the second step with the Metropolis one. This sampling points out the freedom in choosing the transition probability at the second step within the delayed rejection method. However, in this simulation the correlation time, reported in Table 2, is higher than the one found when both steps use Langevin sampling. This result evidences the interplay between $T_1$ and $T_2$ in Eq. 5 in determining the overall acceptance: the Langevin $T_1$ is lower than the Metropolis $T_1$, decreasing the probability of the second move to be accepted.

The acceptance ratio and the mean accepted displacement as function of the distance from the nucleus for the delayed rejection Metropolis algorithm are reported in Figures 2 and 3. The average acceptance ratios for the two moves with $\tau_1 = 0.07$ hartree$^{-1}$ and $\tau_2 = 0.005$ hartree$^{-1}$ are respectively 50% and 66%. In the same Figures the results for two simulations with standard Metropolis algorithm moving one electron at a time and time steps $\tau = 0.07$ hartree$^{-1}$ and $\tau = 0.005$ hartree$^{-1}$ respectively are also reported for comparison. The acceptance ratio for $\tau = 0.07$ hartree$^{-1}$ obviously overlaps the acceptance ratio of the first time step of the delayed rejection algorithm. It is close to zero in the region near the nucleus, then it rises until reaching a constant value of about 60%. The acceptance ratio for the second time step $\tau = 0.005$ hartree$^{-1}$ of the delayed rejection algorithm overlaps the acceptance ratio of the standard Metropolis algorithm with the same time step near the nucleus, while at larger distances it stabilizes at lower values, anyway around 80%, as the acceptance is defined by Eq. 5 instead of Eq. 3. Overall, the acceptance ratio is different from zero also in the core region. The effect of the delayed rejection algorithm is better shown by the average displacement $\langle \Delta R \rangle$ as function of the distance from the nucleus (see Fig. 3). During the first step $\langle \Delta R \rangle$ is relevant only in the valence region and goes to zero near the nucleus. The second step causes the electron to move in the core region, but it slightly affects also the displacement in the valence region. In fact, only in few cases electrons try the second move, because of the high acceptance ratio at the first step. So the average displacement is affected mainly by the first time step in the valence region and by the second one in the core region. Overall, the global average displacement is larger than the $\langle \Delta R \rangle$ of the standard Metropolis algorithm at every distance from the nucleus.

Figures 4 and 5 are similar to Figures 2 and 3, but now the algorithm is the delayed rejection Langevin. The time steps chosen in order to minimize the energy correlation time are 0.07 and 0.003 hartree$^{-1}$ respectively for the first and the second step, while the global acceptance ratios are 82% and 49%. Again as before the electron displacement is dominated in the core by the smaller time step move and in the valence by the larger one. Overall, $\langle \Delta R \rangle$ is larger than the displacement of the standard Langevin sampling at any distance from the nucleus.

Thanks to the improved efficiency in sampling the whole space, ergodicity is guaranteed also in the core region. This is an important result, since a better local sampling allows a greater confidence in evaluating properties largely dependent on the electron density in the core region.

Obviously, the delayed rejection algorithm, despite the improvement in the autocorrelation time, requires more CPU time with respect to the standard algorithms: this is due to the fact that for each rejected step another move is tried, causing a new evaluation of the wave function, its gradient and its Laplacian. The time needed for a delayed rejection simulation is between 20% and 40% longer than the time of a standard one electron at a time simulation. Nevertheless, this drawback is more than rewarded by the improvement in the autocorrelation time and by the better sampling of the core region.

A comparison of our results on Ne with previously proposed acceleration methods[13-15] is not easy as different wave functions were adopted as well as different time steps. Umrigar by his spherical polar coordinate directed Metropolis method[14] got a significant reduction of the correlation time, but he did not perform a "microscopic" analysis, so it is difficult to judge how much the improvement depends on a correct sampling of the core region. Sun *et al.* method[13] slightly improves the mean accepted displacement at short distances from the nucleus, but their best time step 0.03 hartree$^{-1}$ allows a rather poor sampling of the valence space. Slightly better values for the mean accepted displacement in the core region are computed by Mella *et al.*[15]. For = 0.07 hartree$^{-1}$ their improved transition matrix gives $T_{corr}$ = 6.65(3), while the corresponding delayed rejection value is 3.5. All these methods improve the sampling with a negligible increase of CPU time, while the computational cost of the delayed rejection algorithm is higher. One might implement the delayed rejection method on one of these acceleration algorithms to further improve their performance, trying to recover the rejected moves.

The delayed rejection algorithm might be also effective in Langevin simulations when the quantum force F(**R**) becomes very large, that is when a walker is near a nodal surface or in atomic cluster simulations as atoms coalesce[23]. Then the attempted move and therefore the transition probability $T_1$(**R'** **R**,ô$_1$) are very large, while $T_1$(**R** **R'**,ô$_1$) is much smaller. In this case the attempted move is very likely to be rejected and the walker gets trapped. Imposing an arbitrary cutoff on F(**R**) might bias the simulation, even if the probability for a walker to move to a position where the value of F(**R**) is extremely large is very low. Instead, a second attempted move with a shorter time step ô$_2$ has the effect of reducing the contribution of the drift on the attempted move, even if, owing to the very hight value F(**R**) might assume, the reduction of the time step should be substantial.

## Conclusions

We have proposed the delayed rejection algorithm to cope with the multiple time scale problem. The second step, after the first one has been rejected, can be realized with a different transition matrix, for example by using a smaller time step, and so allowing the motion of the core electrons almost as efficiently as the valence electrons. Applications of this method to Metropolis and Langevin sampling of Be and Ne atoms succeeded in improving the sampling of the core region, resulting in a reduction of the correlation time. This method is completely general and can be used with any Monte Carlo application that requires acceleration of the sampling algorithm.

## Acknowledgments

This work has been done in the framework of a project financed by the University of Insubria.

TABLE 1. Time to decorrelate moves for Be with various algorithms. $\tau$ in hartree$^{-1}$, while correlation time is dimensionless.

| Algorithm | $\tau_1$ | $\tau_2$ | Correlation time |
|---|---|---|---|
| Metropolis | 0.03 | | 20 |
| Metropolis: individual electron moves | 0.1 | | 15 |
| Metropolis with delayed rejection | 0.07 | 0.01 | 9 |
| Langevin | 0.07 | | 8 |
| Langevin: individual electron moves | 0.1 | | 7 |
| Langevin with delayed rejection | 0.1 | 0.03 | 5 |

TABLE 2. Time to decorrelate moves for Ne with various algorithms. $\tau$ in hartree$^{-1}$, while correlation time is dimensionless.

| Algorithm | $\tau_1$ | $\tau_2$ | Correlation time |
|---|---|---|---|
| Metropolis | 0.003 | | 50 |
| Metropolis: individual electron moves | 0.05 | | 10 |
| Metropolis with delayed rejection | 0.12 | 0.005 | 5.5 |
| Langevin | 0.01 | | 25 |
| Langevin: individual electron moves | 0.03 | | 7 |
| Langevin with delayed rejection | 0.07 | 0.003 | 3.5 |
| Langevin first move / Metropolis second move | 0.07 | 0.005 | 4.5 |

# Captions

FIG. 1. Number of attempted and accepted moves (arbitrary scale) for Ne as function of the distance from the nucleus. The algorithm is Delayed Rejection Langevin with $\tau_1$=0.03 hartree$^{-1}$ and $\tau_2$=0.007 hartree$^{-1}$ moving one electron at a time.

FIG. 2. The acceptance ratio for Ne as function of the distance from the nucleus. The algorithms are standard Metropolis (M) and Delayed Rejection Metropolis (DR).

FIG. 3. The mean accepted displacement for Ne as function of the distance from the nucleus. The algorithms are standard Metropolis (M) and Delayed Rejection Metropolis (DR) moving one electron at a time.

FIG. 4. The acceptance ratio for Ne as function of the distance from the nucleus. The algorithm is Delayed Rejection (DR) Langevin moving one electron at a time.

FIG. 5. Mean accepted displacements for Ne as function of the distance from the nucleus. The algorithm is Delayed Rejection (DR) Langevin moving one electron at a time.

Bressanini et al.    Figure 1

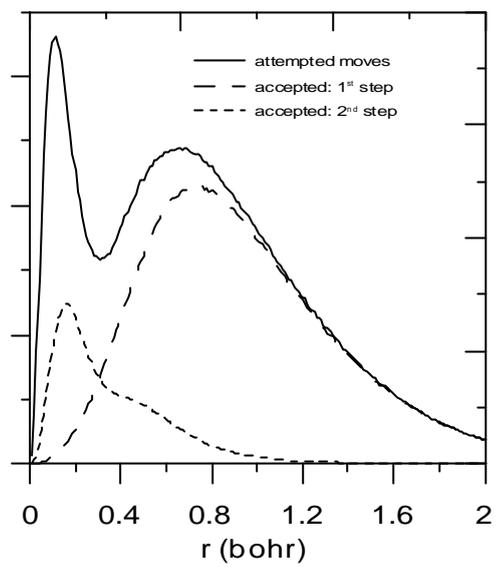

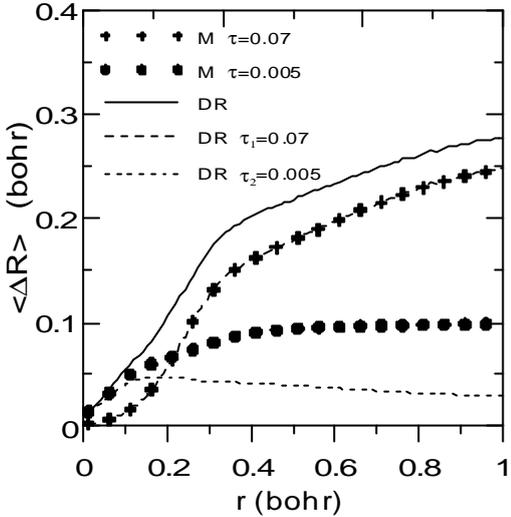

Bressanini et al.     Figure 3

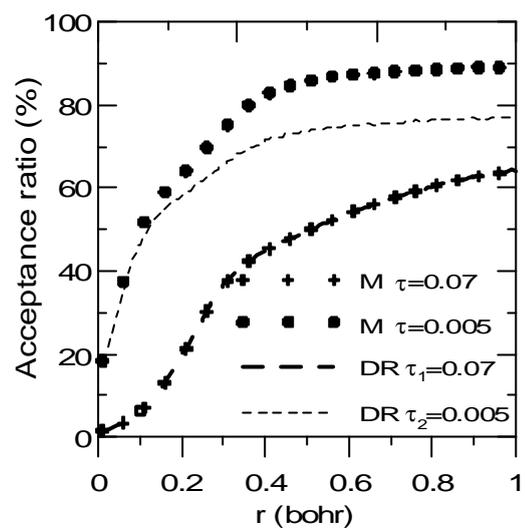

Bressanini et al.    Figure 4

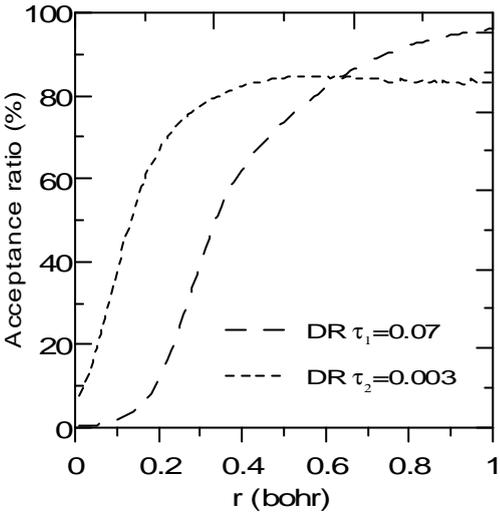

Bressanini et al.    Figure 5

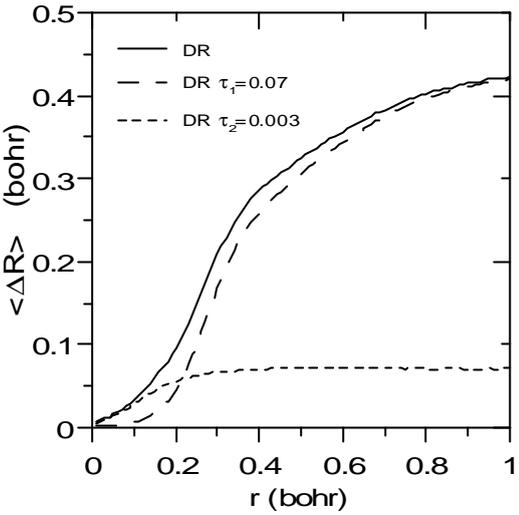